\documentclass[preprint,12pt,3p,sort&compress]{elsarticle}
\usepackage{graphicx}% Include figure files
\usepackage{dcolumn}% Align table columns on decimal point
\usepackage{bm}% bold math
\usepackage{array}
\usepackage{float}

\newcolumntype{C}[1]{>{\centering}m{#1}}
\usepackage{amsmath} % required for `\bmatrix' (yatex added)
\usepackage{fancyhdr}
\usepackage{textcomp}

\journal{Carbon}

\graphicspath{{./images/}}

\begin{document}

\begin{frontmatter}

\title{Graphene and MoS$_2$ interacting with water: a comparison by \textit{ab initio} calculations}

\author[label1]{Giacomo Levita\corref{cor1}}
\address[label1]{CNR-Institute of Nanoscience, S3 Center, Via Campi 213/A, 41125 Modena, Italy}
\ead{glevita@units.it}
\cortext[cor1]{Corresponding author}

\author[label2]{Paolo Restuccia}
\address[label2]{Dipartimento di Scienze Fisiche, Informatiche e Matematiche, Universit\`a di Modena e Reggio Emilia, Via Campi 213/A, 41125 Modena, Italy}

\author[label1,label2]{M. C. Righi\corref{cor2}}
\ead{mcrighi@unimore.it}
\cortext[cor2]{Corresponding author}

\begin{abstract}
Although very similar in many technological applications, graphene and MoS$_2$ bear significant differences if exposed to humid environments. As an example, lubrication properties of graphene are reported to improve while those of MoS$_2$ to deteriorate: it is unclear whether this is due to oxidation from disulfide to oxide or to water adsorption on the sliding surface. By means of \textit{ab initio} calculations we show here that these two layered materials have similar adsorption energies for water on the basal planes. They both tend to avoid water intercalation between their layers and to display only mild reactivity of defects located on the basal plane. It is along the edges where marked differences arise: graphene edges are more reactive at the point that they immediately prompt water splitting. MoS$_2$ edges are more stable and consequently water adsorption is much less favoured than in graphene. We also show that water-driven oxidation of MoS$_2$ layers is unfavoured with respect to adsorption.
\end{abstract}

\end{frontmatter}

\section{\label{sec:intro}Introduction}

Graphene and molybdenum disulfide (MoS$_2$) have recently become some of the most studied nano-materials due to their different array of technological and industrial applications. They are increasingly used as optical and electronic devices, as solid lubricants and as catalytic surfaces for hydrogen storage.\cite{NovoselovRev,GraphTransistor,GraphElect,Erdemir_grafdry,MoS2Review,MoS2Transistor,Restuccia16} These capabilities are related to a structure consisting of layers held together by weak interlayer forces:\cite{Chhowalla13} the layers display at the same time large surface areas and high mechanical resistivity, but also electronic characteristics ranging from semi-metallicity in graphene (a zero gap material) to semi-conductivity MoS$_2$ whose band gap depends on the stacking order.\cite{He14}

However, tribologic and electronic properties are very sensitive to the environment:\cite{Martin96,Qiu12,PRLCle13} as an example humidity induces a very low friction in the sliding motion of graphene sheets\cite{Rietsch13,Scharf13,Carpick10,Erdemir_grafhum} while it hampers the tribologic performances of MoS$_2$.\cite{Vecchio,Nuovo} It is therefore crucial to determine similarities or differences in the reactivity of the two materials towards water in order to understand the microscopic mechanisms underlying such different behaviour and therefore to determine which are the most suited conditions for a determined application.

So far the hydrophilic/hydrophobic behaviour of graphene and MoS$_2$ has been discussed without coming to a clear convergence on the real character of the two materials. The latter in particular has been alternatively described as both hydrophilic\cite{Late12} and hydrophobic;\cite{Gaur14} moreover, its loss of lubricity in humid environments has been interpreted alternatively as consequence of water adsorption on the MoS$_2$ layers\cite{Khare14} or as the result of water-driven oxidation of the material.\cite{Krick11} Such oxidation is considered to be negligible in graphene while in MoS$_2$ it should lead to formation of molybdenum trioxide (MoO$_3$) which could explain the loss of lubricant properties.\cite{Martin07,Sinnott08,Liang11} However, recent experiments have questioned whether this mechanism is effective at room temperature.\cite{Sawyer11,Khare13,Khare14}

Moreover, while the interaction of graphene ribbons with water has been widely studied in the past,\cite{EdgeGraph1,EdgeGraph2} fewer data is available for the analogous interaction between water and MoS$_2$ ribbons.\cite{Shi09,Ghuman15} Edges and vacancies are very sensitive locations for molecular adsorption due to the under-coordination of the atoms on the edge or around the vacancy;\cite{Nardelli05,Susi14,Ciraci11,Ataca11} they also play a special role either in determining the geometrical conformation of layered materials (very reactive edges could disrupt the planar arrangement and lead to irregular or interconnected layers) and in inducing modifications of the electronic and magnetic properties of the layer itself,\cite{Bollinger01,Li08,Sanvito13,Tang14} thus opening the way to interesting possibilities of tuning properties such as the band gap by means of introducing controlled amounts of humidity into the working environment.

By means of static \textit{ab initio} calculations we will perform a comparative study of graphene and MoS$_2$ layers first by investigating water adsorption on the basal planes in order to determine any difference in hydrophilic character. With the same methodology, we will also investigate on the different reactivity of selected defects, such as dislocation and vacancies, and of some stable edges. In the case of MoS$_2$ one possible oxidation mechanism will be discussed and compared to water adsorption. This allows suggesting possible atomistic mechanisms underlying the different macroscopic behaviour of the materials exposed to humidity.

\section{\label{sec:compmet}Computational methods}
The study was carried out by means of Density Functional Theory (DFT) calculations based on plane-wave and pseudopotential expansion of the wavefunction describing the system, as implemented in the Quantum ESPRESSO package.\cite{Espresso} The General Gradient Approximation (GGA) was used to describe the exchange-correlation functional: in particular, we used a Perdew-Burke-Ernzerhof parametrization\cite{PBE,Vanderbilt90} corrected by the semi-empirical Grimme scheme (PBE-D).\cite{Grimme,DFT-D2} The inclusion of van der Waals interactions is necessary when computing water adsorption on bilayers such as graphene and MoS$_2$. The specific PBE-D scheme proved to afford reliable results both in our previous publications\cite{noi14} and in other works on layered materials.\cite{Voloshina11,Silvestrelli11,vdWLayer,Ciraci12,Liu12} The scaling parameter of 0.75 for MoS$_2$ and of 0.65 for graphene were chosen as they correctly reproduced the experimental interlayer distances and binding energies.\cite{noi15} Moreover, they also correct reproduce the geometries and energies obtained with similar calculation methods for water adsorption on the two species\cite{EdgeGraph1,EdgeGraph2,Ghuman15}.

Supercells were used to mimic the basal planes: they were made up of $4 \times 4$ hexagonal elementary cells for MoS$_2$ (including 16 MoS$_2$ units) and of $5 \times 5$ cells for graphene (including 50 C atoms). The choice was suggested by the 1.29 ratio between the lattice constants of graphene and MoS$_2$ (2.47 \r{A} and 3.19 \r{A} respectively). In this way, the two supercells span roughly the same basal plane area (132 \r{A}$^2$ in graphene and 141 \r{A}$^2$ in MoS$_2$). In the case of mono- or bi-layers, the structure was allowed to extend infinitely in the \textit{xy} plane, while periodic replicas along the \textit{z} direction are separated by at least 18 \r{A} of vacuum. When edges were modelled, the supercell was enlarged along one basal direction: this allowed to design ribbons having a width of about 9 \r{A} and separated by 11 \r{A} of vacuum between replicated images along \textit{y}. A $3 \times 3$ \textit{k}-points Monkhorst-Pack grid was used to sample the MoS$_2$ mono- and bi-layers on their \textit{xy} basal plane; for edges the sampling was changed to $4 \times 2$. The equivalent grids employed for graphene were $3 \times 3$ and $6 \times 4$ respectively.

After tests performed on the bulk structures, the kinetic energy cut-off of the plane waves was set to 40 Rydberg. In all calculations, a Methfessel-Paxton smearing\cite{MethPaxt} (0.01 Rydberg for MoS$_2$ and 0.02 Rydberg for graphene) was employed to ease the optimisation procedure and to take into account possible metallization along the edges. Furthermore, edges and defects were investigated also by means of spin-polarized calculation, to determine whether magnetization effects occurred: however, we found that only zig-zag edges proved to be magnetic both in graphene and MoS$_2$. In the Results section we will report energetic data after magnetization was taken into account.

The adsorption energy is obtained as the difference between the total energy of the interacting system and those of the separate substrate and molecule after optimisation within the same calculation cell which is large enough to consider the molecules as isolated. In the case where comparison between adsorption outside or inside a bilayer was carried out, we normalized the total energy difference by the lateral area of the layers. The edge formation energy was evaluated as 1/2*[(E$_{edge} -$ E$_{layer}$)/units-per-edge], where E$_{edge}$ is the total energy of the ribbon and E$_{layer}$ is the total energy of the equivalent structure with the cell shaped as to reproduce an infinite layer. The 1/2 factor takes into account that two edges per ribbon are present in the cell, each of which is made up by the number of units reported as the units-per-edge normalization factor in the formula.

The vacancy formation energy E$_{form}$ on graphene was calculated, like in previous theoretical papers\cite{Nardelli05}, as E$_{form}$ = E$_{vac} -$ (N$-1$/N)*E$_{reg}$ where E$_{vac}$ is the energy of a graphene sheet with a C vacancy, E$_{reg}$ that of the regular sheet and N is the number of C atoms in the calculation cell. On the other hand, for MoS$_2$ E$_{form}$ is obtained as E$_{form}$ = E$_{vac} +$ E$_S -$ E$_{reg}$ where E$_S$ is the energy of a S atom isolated in the vacuum.\cite{Vacancy}

\section{\label{sec:results}Results and discussion}
Graphene and MoS$_2$ are made up of layers with either sulfur or carbon atoms on their basal plane. While graphene is perfectly monodimensional, MoS$_2$ layers are about 3.1 \r{A} wide as the molybdenum plane is sandwiched between two external planes of S atoms. The difference in geometry and coordination between atoms at the centre of the layer and those along their edges is reflected by the markedly different reactivity on such sites. We have therefore analyzed separately the effect of water on the basal plane and on different types of edges.

In the first case, we have considered both undefected and defected monolayers; moreover, bilayer adsorptions were investigated as MoS$_2$ and graphene easily form layered structures held together by (relatively weak) van der Waals interactions. In the second case, we have focused on the standard armchair and zig-zag edges which, as it will be shown, are the most reactive, and on a reconstructed zig-zag edge which according to the literature is the least reactive termination for both MoS$_2$ and graphene.\cite{EdgeGraph1,Spirko03,Helveg11}

\subsection{\label{sec:results-1}Water adsorption on regular and defective layers}
\begin{figure}[H]
  \centering
  \includegraphics[width=0.5\linewidth]{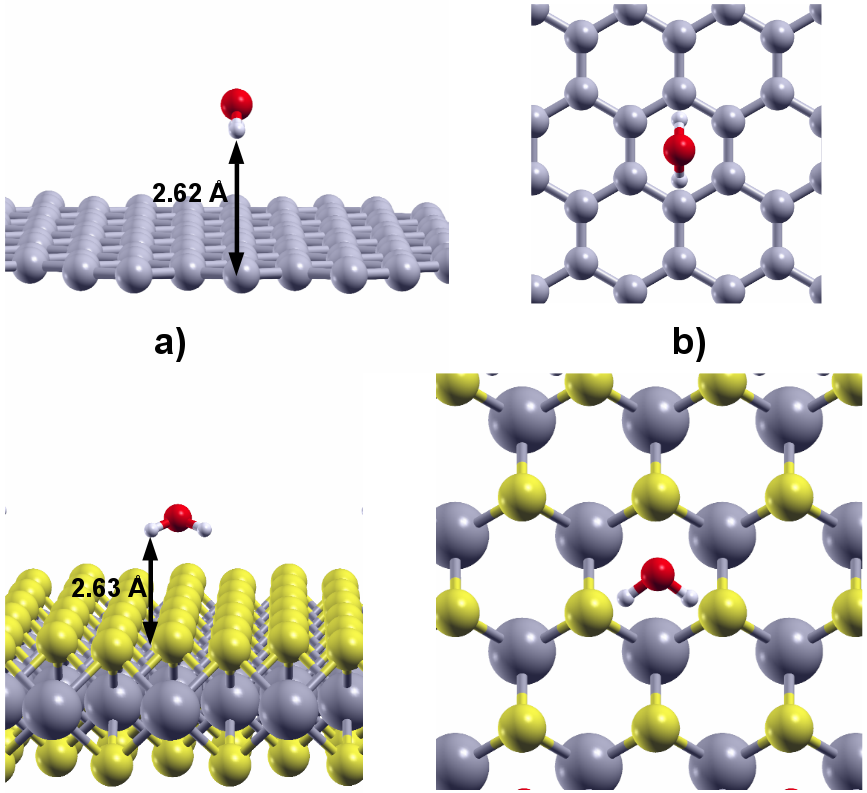}
  \caption{Panel a): side views of water adsorbed on a graphene (above) and a MoS$_2$ monolayer (below). Panel b): top views of the same configurations. Distances between the hydrogen atoms and the surface are reported in \r{A}.}\label{fig:fig1}
 \end{figure}

Due to the lattice structure of both materials, no dangling bonds are present along the external surfaces of the layers of graphene and MoS$_2$, which therefore show a reduced chemical activity. Only physisorption interactions can arise when adsorbing external molecules such as water on this type of structure: this is evidenced by the adsorption energies we calculated for isolated H$_2$O physisorption on monolayer graphene (0.12 eV) and MoS$_2$ (0.15 eV). The resulting geometries are shown in Fig.~\ref{fig:fig1}, along with the distance of the water molecule above the layers. The slightly stronger interaction with MoS$_2$ compared to graphene is due to the larger dimension and polarizability of Mo/S atoms. However, this does not allow to infer any marked difference in hydrophilic character between the two species.

\begin{figure}[H]
  \centering
  \includegraphics[width=0.5\linewidth]{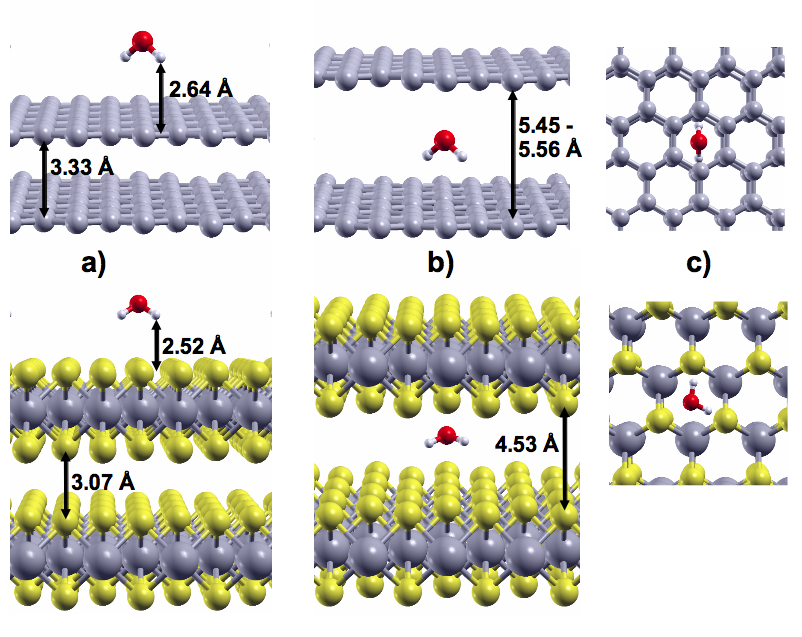}
  \caption{Panel a): side views of water adsorbed above graphene (upper) and MoS$_2$ bilayers (lower). Panel b):  side views of water intercalated between the bilayers. Panel c): top views of the lateral configuration of the bilayers in the intercalated case. Distances are reported in \r{A}.}\label{fig:fig2}
\end{figure}

When a bilayer is formed, as shown in Fig.~\ref{fig:fig2}a, physisorption interactions outside the structure are slightly reinforced as consequence of cooperative effects: the resulting water adsorption energies therefore increase to 0.13 and 0.16 eV for graphene and MoS$_2$ respectively, without any relevant change in distance and orientation of the water molecule above the bilayer with respect to the monolayer. When no water is present, the formation energy of the bilayer has been calculated as 0.23 J/m$^2$ for graphene and 0.27 J/m$^2$ for MoS$_2$; again, MoS$_2$ displays slightly stronger interlayer interactions than graphene albeit of a similar order of magnitude.

If water is adsorbed as an intercalated molecule rather than externally, its presence will keep the layers at larger distances, as shown in Fig.~\ref{fig:fig2}b. This lowers the interlayer binding energy: for a water coverage of about 1 molecule per 140 \r{A}$^2$ it results into a destabilization of 0.17-0.18 J/m$^2$ in both systems. In the case of the more flexible graphene, the effect is evidenced by the slight curvature of the layers around the intercalated molecule: in that region the interlayer distance is therefore 0.1 \r{A} larger than in the areas where no water is present. A further reason (although less effective) for the instability lies in the unfavoured A-A stacking resulting from the necessity of accommodating the intercalated water molecule in the most stable configuration (see Fig.~\ref{fig:fig2}c).

Clearly, such instability could be reduced at higher water coverages, as an increase in water-graphene (-MoS$_2$) interactions could partially compensate the reduced interlayer interactions; our aim here is not to determine such trend but rather to highlight how both graphene and MoS$_2$ bilayers show almost identical behaviour, both geometrically and energetically, when accommodating water within them. Again, from our calculations no major differences can be found in their hydrophilic character. This is highlighted in Table \ref{tab:adsoren}, which reports all water adsorption energies on graphene and MoS$_2$ layers, both undefected and with their most frequent basal defects, which will be discussed just afterwards. We note that while adsorptions on a single layer can be estimated in eV per adsorbate, adsorptions on systems where interlayer interactions are present should be normalized by the area; moreover, the value calculated here for intercalated water only refers to a water coverage of one molecule per 140 \r{A}$^2$ and may be different at other coverages.

\begin{table}[H]
\caption{Adsorption energies for water molecules on graphene and MoS$_2$ monolayers (ML), either regular (reg.) or defective (def.), and bilayers (BL), either with external (ext.) or intercalated (int.) water. Bilayer energies require area normalization and are reported in J/m$^2$; adsorptions on monolayers are evaluated in eV per adsorbate.}\label{tab:adsoren}
\begin{center}
\begin{tabular}{c c c c c c}
\hline
\hline
& reg. ML & def. ML* & BL form. energy & BL ext. & BL int. \\
& (eV) & (eV) & (J/m$^2$) & (J/m$^2$) & (J/m$^2$) \\
\hline
graphene  & -0.12  & -0.22  & -0.23  & -0.016  & +0.167\\
MoS$_2$   & -0.15  & -0.24  & -0.27  & -0.018  & +0.155\\
\hline
\hline
\end{tabular}
\end{center}
\begin{flushleft}
*Stone-Wales defect in graphene; single S-vacancy in MoS$_2$
\end{flushleft}
\end{table}

Literature data proved that Stone-Wales defects and single S-vacancies are the most stable defects on respectively graphene and MoS$_2$ monolayers\cite{Susi14,Vacancy}: therefore we focused on adsorption of water on such basal defects. Our calculated formation energy for a Stone-Wales (SW) defect was evaluated in 5.4 eV; a single C vacancy has instead a formation energy of 7.9 eV and therefore it will not be considered here. The formation energy of the S-vacancy (V$_S$) in MoS$_2$ is calculated as 6.6 eV, similar to the 6.9 eV reported in the literature with similar computational methods\cite{Vacancy2}: the value so obtained exceeds by about 4 eV that obtained when taking into account the chemical potential and the composition of the environment.\cite{Vacancy,Vacancy3} This overestimate is not far from the standard formation enthalpy for MoS$_2$ which is about 2.8 eV.

\begin{figure}[H]
  \centering
  \includegraphics [width=0.5\linewidth] {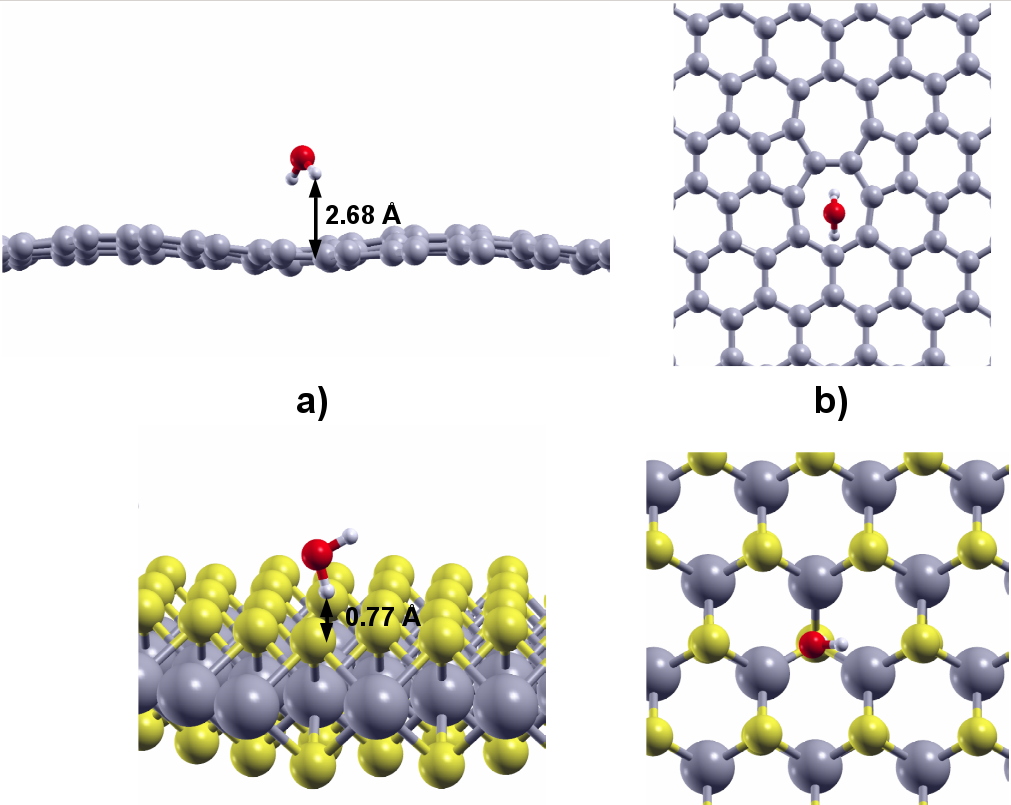}
  \caption{Panel a): side views of water adsorbed above the most stable layer defects of graphene (i.e.: Stone-Wales defect, above) and MoS$_2$ (i.e.: single S-vacancy, below). Panel b): top views of the same configurations. Distances are reported in \r{A}.}\label{fig:fig3}
\end{figure}

Water adsorption on the SW and V$_S$ defects proved to be relatively similar, with energies $-0.22$ eV for SW-graphene and $-0.24$ eV for V$_S$-MoS$_2$ again indicating quite weak physisorption on both systems. The associated geometries are reported in Fig.~\ref{fig:fig3} which for MoS$_2$ shows a more vertical arrangement of the water molecule compared to the non defective case. While MoS$_2$ is a relatively stiff layer and the presence of the defect brings the hydrogen atoms of H$_2$O much closer to the surface, graphene tends to markedly bend outwards below the water molecule (see Fig.~\ref{fig:fig3}a), showing little interaction with it. As a result, the water-carbon distance remains unchanged with respect to the undefected case, whereas in V$_S$-MoS$_2$ it is reduced by about 1 \r{A}.

A further indication of the weak reactivity of both defects towards humid environments is evidenced by the dissociative H/OH adsorption which - similarly to what reported in previous studies on MoS$_2$\cite{Ciraci12a} - is unfavoured in both materials, although by a different amount (+1.36 eV for SW-graphene, +0.35 eV for V$_S$-MoS$_2$). This allows to conclude that even when defected both graphene and MoS$_2$ display limited adsorption of isolated water molecules on their basal plane; such situation could change at higher water coverages (where additional H-bonds stabilize adsorbed molecules or fragments) or by more reactive defects such as C-vacancies in graphene or S$_2$-vacancies in MoS$_2$\cite{Ciraci12a}. However those defects are calculated as highly unstable and less likely to be formed, therefore they will not be discussed here.

\subsection{Water adsorption on lateral edges}
The most important and common defects in the ordered structure of graphene and MoS$_2$ layers are the edge terminations, which present unsaturated atoms with marked reactivity compared to the rest of the structure. We therefore investigated the effect of water on a selection of these edges, considering different adsorption geometries and fragments (i.e. molecular water, H, OH and O fragments, plus oxygen substitution).

\begin{figure}[H]
  \centering
  \includegraphics [width=0.5\linewidth] {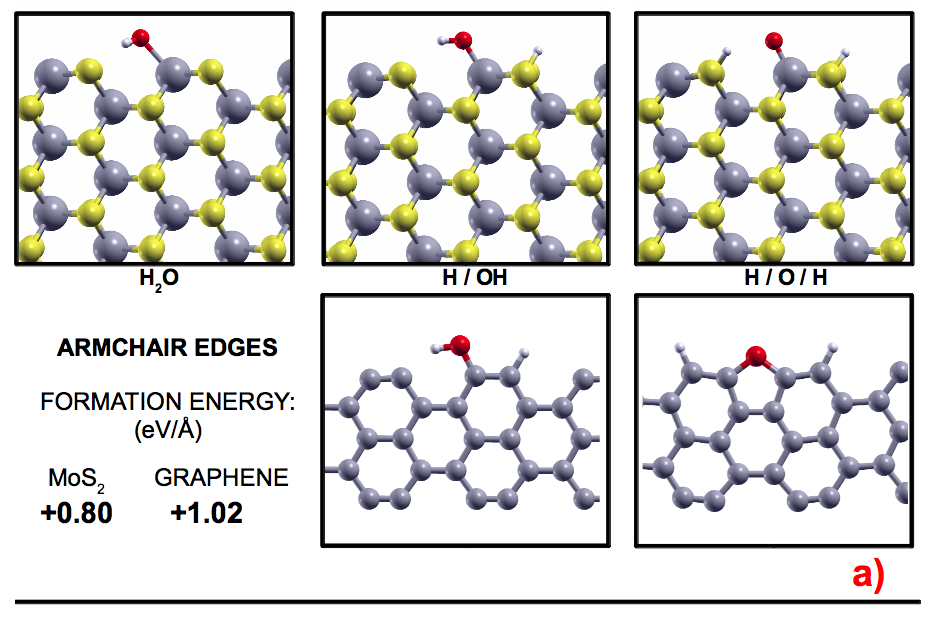}
  \includegraphics [width=0.5\linewidth] {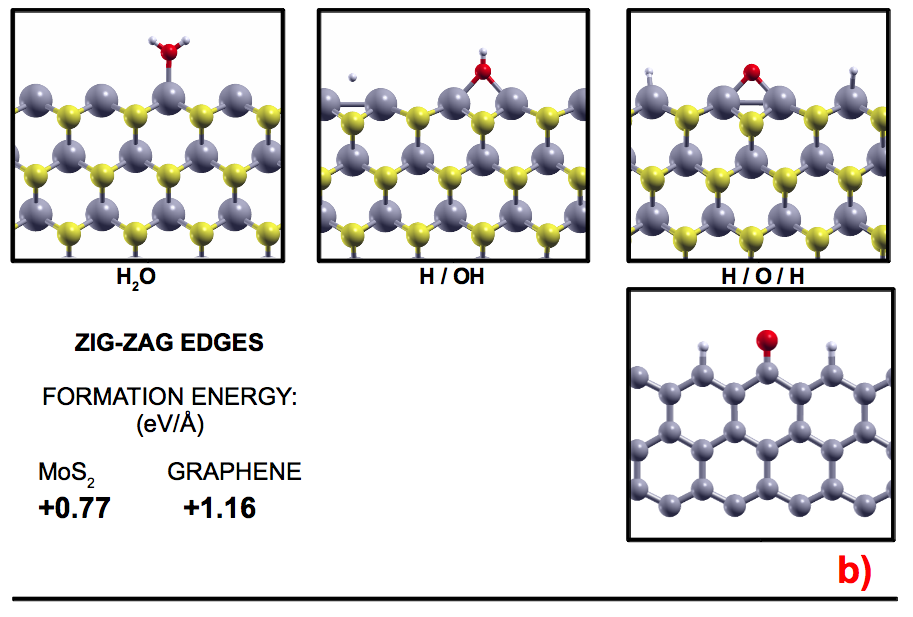}
  \includegraphics [width=0.5\linewidth] {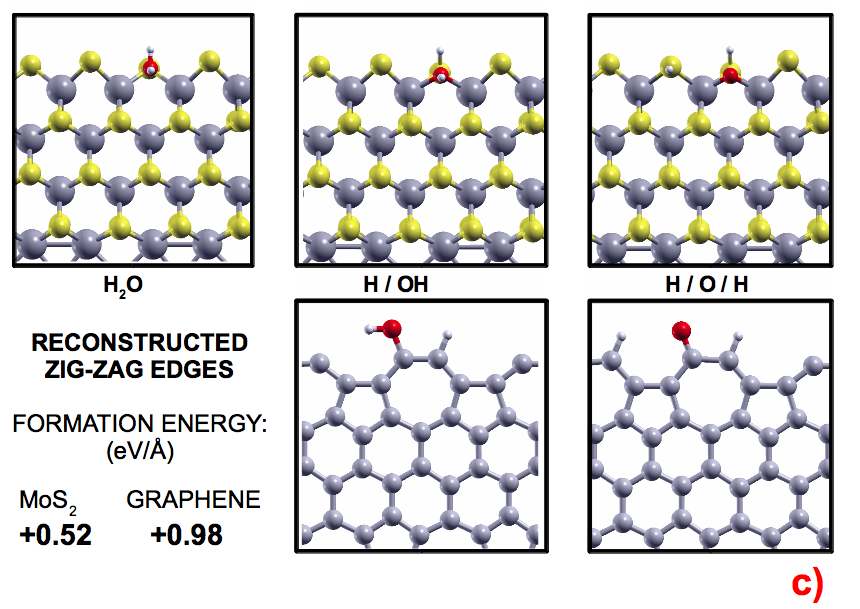}
  \caption{Panel a): top views of the adsorption geometries for water and its fragments on armchair edges. Left column: water on MoS$_2$. Centre column: H/ OH on MoS$_2$ (above) and graphene (below). Right column: H/ O/ H on MoS$_2$ (above) and graphene (below). Panel b): top views of the adsorption geometries for water and its fragments on zig-zag edges. Left: water on MoS$_2$. Centre: H/ OH on MoS$_2$. Right: H/ O/ H on MoS$_2$ (above) and graphene (below). Panel c): top views of the adsorption geometries for water and its fragments on the most stable reconstructed edges. Left: water on S-reconstructed zig-zag MoS$_2$. Centre: H/ OH on MoS$_2$ (above) and 5-7 reconstructed zig-zag graphene (below). Right: H/ O/ H on MoS$_2$ (above) and graphene (below).}\label{fig:fig4}
\end{figure}

We carried out the analysis by first identifying the edge with higher formation energy, as this will be the most reactive towards adsorbates. The evaluation of the formation energy was carried out by means of the formula discussed in the Computational Methods section. Formation energies for each edge are reported in the correpsonding panel of Fig.~\ref{fig:fig4}, which displays the geometries for molecular and dissociative adsorptions on the individual type of edge (respectively armchair in panel a, zig-zag in panel b and a reconstructed zig-zag in panel c).

In the case of MoS$_2$, armchair and zig-zag edges have almost identical formation energy (0.80 eV/\r{A} and 0.77 eV/\r{A}); on the contrary, in graphene the former is more stable (1.02 eV/\r{A} vs. 1.16 eV/\r{A}). More remarkably, the average edge formation energy in graphene is about 0.3 eV/\r{A} higher than in MoS$_2$: this is due to the fact that even in the least stable zig-zag edge the Mo atoms, although undercoordinated, are still bound to four S atoms while the corresponding carbon atoms on graphene edges are only coordinated to two C atoms. Such effect is also evident when considering the most stable edge reconstruction in the two systems (Fig.~\ref{fig:fig4}c): for graphene this is a 5-7 edge reconstruction which halves the number of undercoordinated C atoms per edge without eliminating them (i.e. the C atoms on the 7-sided rings are still undercoordinated). In MoS$_2$ the most stable edge is a reconstructed zig-zag where one row of S atoms sits on the outside of the row of exposed Mo atoms: this allows the formation of Mo$-$S$-$Mo bridges on both edges and therefore a much reduced instability of the metal atoms. This is reflected by its 0.52 eV/\r{A} formation energy, which is far lower than the 0.98 eV/\r{A} for the 5-7 graphene edge.

Contrary to what happened in the case of basal plane defects, edges in graphene prove to be not only less stable than in MoS$_2$ but also much more reactive towards water adsorption: none of the three edges considered here was in fact able to allow molecular H$_2$O physisorption but prompted instead its immediate dissociation. This is a clear indication of a high chemical activity which is not evidenced by MoS$_2$, where molecular water can be absorbed with energies of $-1.34$, $-1.00$ and $-0.43$ eV on the armchair, zig-zag and reconstructed zig-zag edges respectively. As seen in Fig.~\ref{fig:fig4}a, the larger adsorption energy for the armchair edge is due to the formation of an additional H-bond interaction between water and an external S atom; on the contrary, the reconstructed zig-zag edge shows low reactivity due to the lack of exposed Mo atoms (see Fig.~\ref{fig:fig4}c) which are the most important sites for oxygen coordination.

Table \ref{tab:edgeen} summarizes all adsorptions of water (or its fragments) on the selected edges studied here. The Table immediately reveals how MoS$_2$ allows molecular adsorption while graphene does not; moreover, adsorptions on MoS$_2$ are always less favoured than on the equivalent graphene edge. We also note that while for graphene our values are almost in perfect agreement with those previously reported\cite{EdgeGraph1,EdgeGraph2}, our water adsorption on MoS$_2$ is somewhat larger than the 0.55 eV reported in the literature\cite{Ghuman15}: we attribute such outcome to the different calculation setup employed here, especially the inclusion of magnetization (and to a smaller extent the different treatment of the van der Waals forces).

\begin{table}[H]
\caption{Molecular and dissociative adsorption energies on armchair, ziz-zag and reconstructed (Reconstr.) edges for graphene or MoS$_2$. The associated geometries are reported in Fig.~\ref{fig:fig4}. Energies are reported in eV per molecule.}\label{tab:edgeen}
\begin{center}
\begin{tabular}{c c c c c c c}
\hline
\hline
&  & Graphene &  &  & MoS$_2$ & \\
& Armchair & Zig-Zag & Reconstr. & Armchair & Zig-Zag & Reconstr.\\
\hline
H$_2$O  & --  & --  & --  & $-1.34$  & $-1.00$ & $-0.43$\\
H/OH  & $-3.87$  & --  & $-2.66$  & $-1.96$  & $-3.36$ & $-0.56$\\
H/H/O  & $-6.28$  & $-7.02$  & $-2.61$  & $-2.20$  & $-3.11$ & $+0.68$\\
\hline
\hline
\end{tabular}
\end{center}
\end{table}

As said, water can be absorbed on graphene edges only upon fragmentation: when OH and H are produced they are absorbed with energy $-3.87$ eV for the armchair edge and $-2.66$ eV for the 5-7 reconstruction. These values are much larger than the $-1.96$ and $-0.56$ eV adsorption energies on the corresponding MoS$_2$ edges. Even more indicative, for graphene the zig-zag edge does not allow even the OH/ H dissociation itself, and a full fragmentation into H/ O/ H immediately occurs; on the contrary, on MoS$_2$ such adsorption is feasible with energy $-3.36$ eV. Such value (larger than the corresponding one on the other edges) is a consequence of the saturation of four external atoms, as the fragments sit in bridge positions between adjacent Mo atoms (see Fig.~\ref{fig:fig4}b).

It is only at the final stage of a H/ O/ H dissociation that a full comparison between the six edges becomes possible. Again, graphene shows large adsorption energies at $-6.28$, $-7.02$ and $-2.61$ eV for the armchair, zig-zag and 5-7 reconstruction respectively. The energy for the 5-7 reconstruction (50 meV lower than in the OH/H adsorption) is due to the instability of a ketonic C=O group compared to the alcoholic C$-$OH group (see Fig. 4c). In MoS$_2$ the corresponding values are $-2.20$, $-3.11$ and $+0.68$ eV which are between 3 and 4 eV higher than in graphene. The positive value for the S-reconstructed edge is due to the difficulty of forming the Mo$-$O$-$Mo bonds which in the zig-zag edge stabilized the structure (Fig.~\ref{fig:fig4}c); also on the zig-zag edge a full dissociation is less favourable than a partial OH/ H one. We therefore infer that only the reactive armchair edge can allow a full H/ O/ H adsorption.

Finally, as in the past it was proposed that for MoS$_2$ water may prompt formation of MoO$_3$ by direct oxidation\cite{Martin07,Liang11}, we checked out also the possibility of substituting a sulfur atom on the edge with the oxygen atom from the water molecule. This leads to the formation of a molecule of hydrosulfuric acid near a Mo$-$O$-$Mo bridge on the exposed edge: however, even for the reactive zig-zag case this arrangement proves to be less stable ($+0.43$ eV) than when water is physisorbed on the regular edge. While this does not necessarily preclude the possibility of an actual oxidation taking place, it is nevertheless much less favourable than a simple adsorption which, as seen, is associated with formation energies of around $-2$ eV. We therefore conclude that at ordinary temperatures MoS$_2$ is more likely to adsorb water along its edges rather than being oxidised by it.

The overall outcome is that graphene edges - whose formation energy is higher than in MoS$_2$ - show a much larger tendency of adsorbing water fragments compared to MoS$_2$ which instead also allows molecular adsorption. These large differences in adsorption energy imply that graphene will easily capture water fragments along its edges while MoS$_2$ in certain conditions could allow water diffusing through the interface between basal planes. This, as shown in Section~\ref{sec:results-1}, is not a favourable arrangement compared to when water is confined outside the bilayer. On the contrary, graphene could easily keep water out of these interfaces by “trapping” them on the reactive edges, as it has already been suggested experimentally by noticing that graphite does not seem to change its interlayer distance in humid environments\cite{APL04}. Moreover, if no water is present the large reactivity of graphene edges can lead to their interconnection: this may result into a “jamming” of graphene layers and eventually to the elimination of extended or ordered monolayers. Humidity could therefore prevent such effect\cite{Rietsch13} and help in maintaining an ordered graphene stacking.

\section{\label{sec:conclus}Conclusions}
The unique properties of graphene and molybdenum disulfide are largely affected by the interaction of their layers with water: humid environments induce relevant differences between these materials and such macroscopic differences reflect different interactions occurring between molecules and surfaces at the nanoscale. We showed here that on MoS$_2$ water molecules do not favour oxidation of the basal planes; we also showed that on both graphene and MoS$_2$ basal planes physisorption occurs with about the same adsorption energy either in their regular or defective arrangement.
Moreover, both systems show a similar trend when water is intercalated between bilayers: such arrangement is unfavoured by approximately the same energetic amount compared to the adsorption outside the bilayer. In other words, the two interlayer interfaces show a comparable hydrophobic character which is mostly due to the weakening of van der Waals interactions within the bilayers.

The only sites where graphene and MoS$_2$ display a well diverging behaviour are the edge terminations of the layer: on both the most and the least stable ones, graphene edges are so reactive that water can not be adsorbed as an individual molecule but is fragmented in either H or OH ions or, in the case of the most reactive zig-zag edge, directly in two H and one O ions. On the contrary, MoS$_2$ edges have lower formation energies: this induces consequently a lower chemical activity which allow them to adsorb molecular water without dissociating it.
Of course chemisorption may occur on MoS$_2$ too: however, also in these cases graphene edges show adsorption energies 3 or 4 eV larger than in MoS$_2$. It is remarkable that the most stable edge termination in MoS$_2$ (i.e.: a sulfur-reconstructed zig-zag edge) does not favour a total H/ O/ H dissociation which in graphene is the most favoured type of adsorbate.

In summary, our calculations show that the mechanisms of water adsorption are similar on the basal planes of the two materials but are largely different along the edges due to their different chemical activity. Such difference may impact the wetting mechanism which has been shown to be strictly related to the tribological properties of layered\cite{Khare14} and carbon based materials\cite{SeijiCarb}.

\section*{\label{sec:ack}Acknowledgements}
We acknowledge the CINECA consortium for the availability of high performance computing resources and support through the ISCRA-B "attrit0" project.

\section*{References}

\bibliographystyle{elsarticle-num}
\bibliography{Edges}

\end{document}